\documentclass[AMA,STIX1COL]{WileyNJD-v2}
\usepackage{moreverb}

\newcommand\BibTeX{{\rmfamily B\kern-.05em \textsc{i\kern-.025em b}\kern-.08em
T\kern-.1667em\lower.7ex\hbox{E}\kern-.125emX}}

\articletype{Case Study}%

\received{<day> <Month>, <year>}
\revised{<day> <Month>, <year>}
\accepted{<day> <Month>, <year>}

\begin{document}

\title{Redrawing the 2012 map of the Maryland congressional districts}%\protect\thanks{This is an example for title footnote.}}

\author[1]{Noah Lee}

\author[2]{Hyunwoo Park}

\author[3]{Sangho Shim*}

\authormark{Lee \textsc{et al}}

\address[1]{%\orgdiv{Org Division}, 
\orgname{North Allegheny Senior High School}, \orgaddress{\state{Pennsylvania}, \country{USA}}\protect}

\address[2]{\orgdiv{Graduate School of Data Science}, \orgname{Seoul National University}, \orgaddress{\state{Seoul}, \country{Republic of Korea}}}

\address[3]{\orgdiv{School of Engineering, Mathematics and Science}, \orgname{Robert Morris University}, \orgaddress{\state{Pennsylvania}, \country{USA}}}

\corres{*Sangho Shim, 6001 University Blvd, Moon Twp, PA 15108, USA. \email{shim@rmu.edu}}

\presentaddress{6001 University Blvd, Moon Twp, PA 15108, USA.}

\abstract[Abstract]{Gerrymandering is the practice of drawing biased electoral maps that manipulate the voter population to gain an advantage. The most recent time gerrymandering became an issue was 2019 when the U.S. Federal Supreme Court decided that the court does not have the authority to dictate how to draw the district map and state legislators are the ones who should come up with an electoral district plan.
We solve the political districting problem and redraw the 2012 map of Maryland congressional districts which raised the issue in 2019.}


\maketitle

\section{Introduction}\label{sec1}

Gerrymandering is the practice of drawing biased electoral maps that manipulate the voter population to gain an advantage. It is an issue that is relatively unknown despite its vast importance and relevance to the modern American political landscape. The maps are drawn in a way that isolates or weakens opposition to the point where it is virtually impossible for an opposing party to have a significant win.  
 
Gerrymandering has been an issue that has plagued democracy in America for nearly its entire history. Many cite Governor Elbridge Gerry as the source of this issue; in March of 1812, Elbridge drew an electoral map of Massachusetts that twisted and turned so drastically that it was called ``a new species of monster...the Gerry mander'' by the Boston Gazette.~\cite{little2021} The nickname was derived from the salamander-like shape of the recently %{\textcolor{blue}{recently}} 
drawn map and Elbridge Gerry’s last name. This map, massively favoring the Jeffersonian Republicans, is memorialized as the origin of the term gerrymandering. 
However, gerrymandering has existed far before this fateful incident. The origins of this issue can be traced to 18th century England, and following the English movement to the New World, gerrymandering began almost immediately in the new colonies. 

Gerrymandering continued to persist and even grew in prevalence following the South’s defeat in the Civil War.~\cite{little2021} African Americans winning their civil rights, namely voting rights, was seen as a massive threat to white dominance in the South. Gerrymandering was employed by Southern politicians as a way of minimizing the impact of African Americans votes; this method of voter suppression eventually fell off in favor of other methods such as the infamous Jim Crow laws. Gerrymandering has still remained throughout all these years and with the aid of new methods has become a mainstay issue in modern politics.
Preventive measures were passed in the 1960s by the Supreme Court. Led by Chief Justice Earl Warren, the Supreme Court ruled that electoral districts must be roughly equal in population and must be redrawn every 10 years based on the census.
 
There are two types of Gerrymandering: cracking and packing. Packing refers to the practice of drawing electoral lines that group people of similar political views into one region.~\cite{kirschenbaum2021} This method weakens its victim's strength in other districts by placing the majority in one single district; critically, a single district win will not win a state. North Carolina’s 12th district was a prime example of this technique; the primary issue of the map was that Republican lawmakers packed Democrat voters into the 12th district while ignoring location.~\cite{bailey2020} The other method is known as cracking and is the separation of voters of similar political views. This causes them to be divided in small pockets, and these pockets do not have the population to win a district. Maryland’s 2012 electoral map was a case of cracking Republican support. One of the most predominantly Republican districts was cracked and spread thin to weaken Republican power.~\cite{bailey2020}
 
This paper redraws the 2012 map of the Maryland congressional districts. Chopra, Park and Shim~\cite{chopra2023} redrew the map partitioning 46 coarse population units into the 8 congressional districts. However, the real-world political districting problem partitions 1,849 precincts in Maryland. We develop a randomized rounding heuristic and solve the large-scale political districting problem.  
Section~\ref{s:problem} introduces the political districting problem which Mehrotra, Johnson and Nemhauser~\cite{mehrotra1998optimization} introduced. In Section~\ref{s:heuristic}, we develop a rounding function to find an integer solution close to a fractional solution, and remake the adapative randomized rounding procedure, which Kim and Shim~\cite{KS2022} introduced, using our rounding function. In Section~\ref{s:case}, the adaptive randomized rounding approach partitions the finest population units of precincts redrawing the 2012 map of the Maryland congressional districts. In Section~\ref{s:conclusion}, we compare the redrawn map with the original 2012 map and discuss computational issues of the political districting problem. Throughout this paper, we use graph theoretical terms following Bondy and Murty~\cite{BM2010}.

\section{Problem formulation}\label{s:problem}
In this section, we develop a mathematical model of the political districting problem which Mehrotra, Johnson and Nemhauser~\cite{mehrotra1998optimization} introduced. The political districting problem designs a district plan which is a partitioning of indivisible population units (for example, counties, precincts, etc.) into a predetermined number of districts such that the units in each districts are contiguous, each district is geographically compact and the sum of the populations of the units in any district lies within a predetermined range. 

The problem can be modeled as a graph partitioning problem by associating a node with every population unit and connecting two nodes by an edge whenever the corresponding population units are geographically neighbors. The weight on a node is equal to the population of the corresponding unit. A plan is represented by a partitioning of the nodes such that the nodes in any set of the partition induce a connected subgraph (to ensure contiguity of the districts) and the sum of the node weights lies within the prespecified range (to satisfy the population requirements). A plan is good if each resulting district is geographically compact.

Following Mehrotra~et~al.~\cite{mehrotra1998optimization}, we assign a penalty cost to every potential district that measures its ``non-compactness" from an ideal district, which we refer to as gerrymander score. The gerrymander score is total distance where the distance function measures the proximity in terms of how many other population units one must go through to get from one population unit to another. If all possible districts that satisfied the population and contiguity requirements were enumerated, the problem would determine a set of $k$ districts, where $k$ is known, that minimizes the gerrymander score while making sure that every population unit is included in exactly one of the selected districts. 

Let $G(V, E)$ be a graph with $V$ defined to be the set of population units and $E$ the pairs of units that share a common border.
We refer to $G(V,E)$ as adjacency graph.
Let $k$ be the number of districts and let $D=\{ 1,2,...,k\}$ denote the set of district numbers. 
The political districting problem partitions the population units $V$ into districts $V_1,V_2,...,V_k$, such that $V = \bigcup_{d\in D} V_d$ and $V_{d_1}\cap V_{d_2}=\emptyset$ for $d_1\neq d_2\in D$. 
In the political districting problem, all the districts induce connected subgraphs $G[V_d],d=1,...,k$.
That is, a district plan is so called a connected partition.
In this paper, the set of connected partitions will be denoted by $C(V,D)$.
We employ binary variables $Y(u,d)\in\{ 0,1\}$ indicating the membership of each node $u\in V$ into district $V_d$; {\it i.e.}, $Y(u,d) = 1$ if node $u$ belongs to district $V_d$, and $Y(u,d) = 0$ otherwise.
The set of $Y$-vectors indicating the connected partitions will be denoted by $C(V,D)$, too.

The sum of the populations of the units in any district is supposed to lie within a predetermined range. 
Let $p_u$ denote the population of the unit $u\in V$, and let the mean population of a district be $\bar{p}={\sum_{u\in V}p_u}/{k}$. Then, the maximum deviation $\delta$ (typically $\delta=0.05$) allowed from the mean population defines the lower and the upper bounds of a district; $p_{\min}=(1-\delta)\bar{p}$, $p_{\max}=(1+\delta)\bar{p}$. The equal population constraints are
\begin{eqnarray*}
p_{\min}\leq \sum_{u\in V} p_u Y(u,d)\leq p_{\max}\mbox{ for }d\in D.
\end{eqnarray*}
We will refer to a plan of maximum deviation $\delta$ as $100\delta\%$-plan.
A connected partition will be said to be feasible if all the equal population constraints are satisfied.

Mehrotra~et~al.~\cite{mehrotra1998optimization} developed a penalty cost that measures noncompactness of a district to get a proxy for the visual notion of non-compactness that is easy to use in an optimization model, noting that a district would tend to be compact if the population units in the district are not far from each other. Consider the node induced subgraph $G^{d} (V^{d}, E^{d})$ of district $d$ that is connected and satisfies population bounds. We can measure the non-compactness of an induced subgraph $G^{d}$ by how far units in the district are from a central unit. The length of a path from $u$ to $v$ is defined to be the number of edges in the path. Let $s_{uv}$ be the number of edges in a shortest path from $u$ to $v$ in $G$. We define the center of $G^{d}$ to be a node $u \in V^{d}$ such that $\sum_{v\in V^{d}} s_{uv}$ is minimized. We define the cost (called gerrymander score) of the district to be $\sum_{v\in V^d} s_{uv}$ where $u$ is a center of the district. The smaller the cost, the more compact a district is. 

The political districting problem is
\begin{eqnarray}
    \min &&S(Y)\label{e:obj}\\
    \mbox{subject to}&&Y\in C(V,d)\label{e:connected}\\
    &&p_{\min} \leq \sum_{u\in V}p_u Y(u,d) \leq p_{\max}\mbox{ for }d\in D\label{e:population}
\end{eqnarray}
where $S(Y)$ is total gerrymander score of the district plan indicated by $Y$. The optimal solution to the model is called the optimal $100\delta$\%-plan.

\section{Adaptive Randomized Rounding}\label{s:heuristic}
Modifying the rounding function which Chopra, Qiu and Shim~\cite{CQS2023} introduced, we develop a rounding function on $[0,1]^{V\times D}\subset \mathbb{R}^{V\times D}$ to identify the nearest integer solution indicating a district plan. We then solve the political districting problem in the framework of adaptive randomized rounding (ARR) procedure which Kim and Shim\cite{KS2022} introduced.
Our ARR procedure tackles the problem (\ref{e:obj})-(\ref{e:connected}) with a general objective function $S(Y)$ to minimize.
The equal population constraint~(\ref{e:population}) will be the objective function $S(Y)$ in Phase~I.
Once a feasible solution $Y$ is found satisfying (\ref{e:population}), the objective function will be total gerrymander score and the infeasible solutions found in the next trials will be ignored. 

Given a fractional solution $\tilde{Y}\in [0,1]^{V\times D}\subset\mathbb{R}^{V\times D}$, which we will refer to as a \emph{seed}, randomized rounding finds an integer solution $Y^{\mathrm{INT}}\in C(V,D)$ near the seed. 
A randomized rounding procedure finds multiple integer solutions from multiple trials and selects the best integer solution found, denoted by $Y^{\mathrm{BEST}}$. 
An adaptive randomized rounding procedure moves the seed $\tilde{Y}$ toward the best known integer solution $Y^{\mathrm{BEST}}$ to find a better integer solution near $Y^{\mathrm{BEST}}$ ({\it i.e.}, a better integer solution in the approximate neighborhood of $Y^{\mathrm{BEST}}$). 

Section~\ref{s:three} describes the three main components of our adaptive randomized rounding technique, and Section~\ref{s:ARR} shows the construction of the full adaptive randomized rounding procedure. Then, Section~\ref{s:phase1} introduces a conditional objective function to identify a feasible solution in Phase~I and to minimize total gerrymander score in Phase~II.

\subsection{Three Main Components}\label{s:three}
\subsubsection{Rounding}
A rounding function $\mathrm{ROUND}:\mathbb{R}^{V\times D}\rightarrow\mathrm{INT}(D)$ maps a fractional solution $\tilde{Y}\in\mathbb{R}^{V\times D}$ to the nearest integer solution $Y^{\mathrm{INT}}=\mathrm{ROUND}\left(\tilde{Y}\right)$ indicating a connected partition ({\it i.e.}, the nodes of each district indicated by $Y^{\mathrm{INT}}$ induce a connected sub-graph of the adjacency graph).
Algorithm~\ref{a:rounding} is a pseudo-code of the rounding function. 
Given a fractional vector $\tilde{Y}=\left( \tilde{Y}(u,d): u\in V,d\in D\right)$, the rounding function identifies the nearest integer solution $Y^{\mathrm{INT}}=\mathrm{ROUND}\left( \tilde{Y}\right)$ indicating a connected partition.
As the usual rounding function of real numbers ({\it e.g.}, $\mathrm{ROUND} (0.1)=\mathrm{ROUND} (0) = 0$ and $\mathrm{ROUND} (0.8)=\mathrm{ROUND} (1) = 1$), an integer solution $Y^{\mathrm{INT}}$ is rounded to itself; {\it i.e.}, 
\begin{eqnarray}
\mathrm{ROUND}\left(Y^{\mathrm{INT}}\right)=Y^{\mathrm{INT}}.\label{e:round}
\end{eqnarray} 

\begin{algorithm}
\caption{Rounding Algorithm}
\label{a:rounding}
\hspace*{\algorithmicindent} \textbf{Input:}  A fractional vector $\tilde{Y}\in\mathbb{R}^{V\times D}$ \\
\hspace*{\algorithmicindent} \textbf{Output:} An integer vector $Y^{\mathrm{INT}}\in\{ 0,1\}^{V\times D}$ indicating a connected partition
\begin{algorithmic}[1]
\State (Initialization of sets) $V_d = \emptyset$ for district $d\in D$; $W=V$ \label{l:initialize}
\State (Initialization of vector) $Y^{\mathrm{INT}} (u,d) = 0$ for $u\in V$ and $d\in D$ \label{l:initialize2}
\While{$W\neq \emptyset$}\label{l:whileStarts}
    \If{$V_d=\emptyset$ for some $d\in D$}\label{l:ifempty}
        \State Identify the node $u\in W$ maximizing $\tilde{Y}(u,d)$.
        \State $V_d \leftarrow \{ u\}$.
        \State $W \leftarrow W\setminus \{ u\}$\label{l:endifempty}
    \Else\label{l:else}
        \State Identify $\left( w^*,d^*\right)\in W\times D$ maximizing $\tilde{Y}\left( w^*,d^*\right)$ with $N\left( w^*\right)\cap V_{d^*}\neq\emptyset$. \Comment{$N\left( w^*\right)$: the neighborhood of $w^*$}
        \State $Y^{\mathrm{INT}}\left( w^*, d^*\right) = 1$\label{l:observe}
        \State $V_{d^*} \leftarrow V_{d^*}\cup\{ w^*\}$; $W \leftarrow W\setminus \{ w^*\}$\label{l:endelse}
    \EndIf
\EndWhile\label{l:whileEnds}
\State Return $Y^{\mathrm{INT}}$ indicating connected partition $\left( V_j:j\in J\right)$
\end{algorithmic}
\end{algorithm}

\begin{figure}
\begin{subfigure}{.5\textwidth}
  \centering
  % include first image
  \includegraphics[width=.8\linewidth]{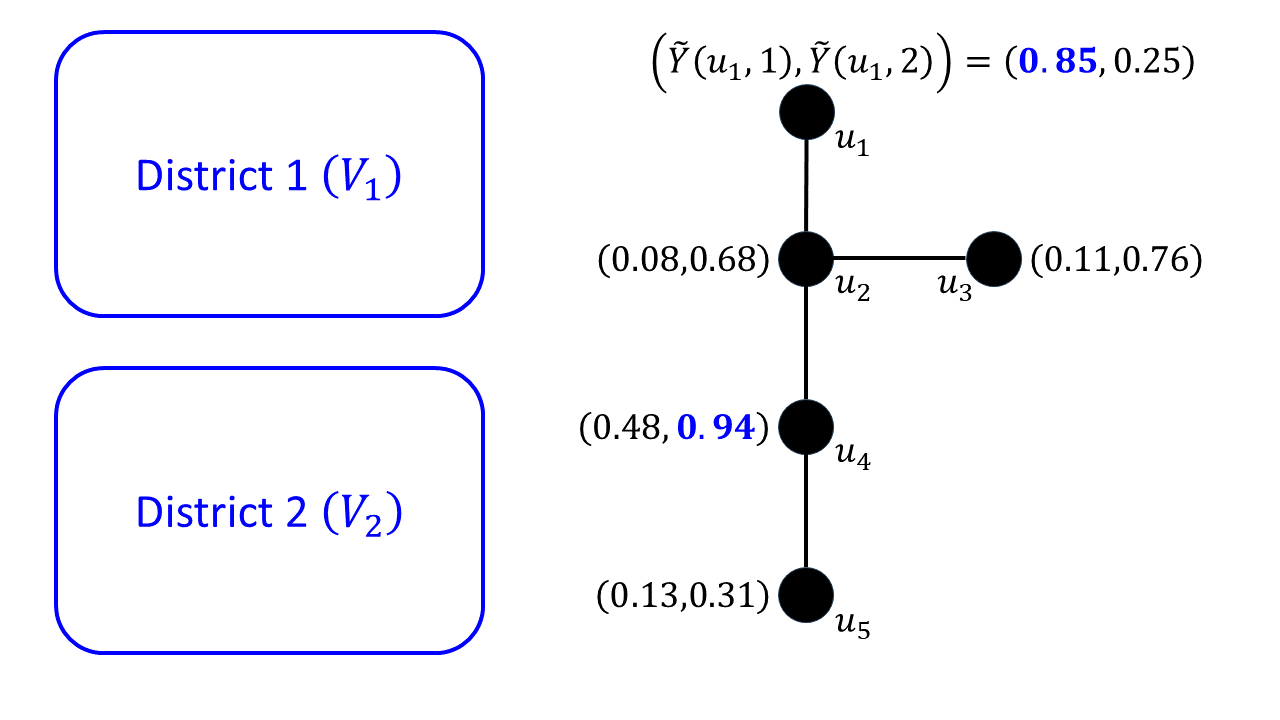}  
  \caption{Put $u_1,u_4$ into empty $V_1,V_2$}
  \label{fig:sub-first}
\end{subfigure}
\begin{subfigure}{.5\textwidth}
  \centering
  % include second image
  \includegraphics[width=.8\linewidth]{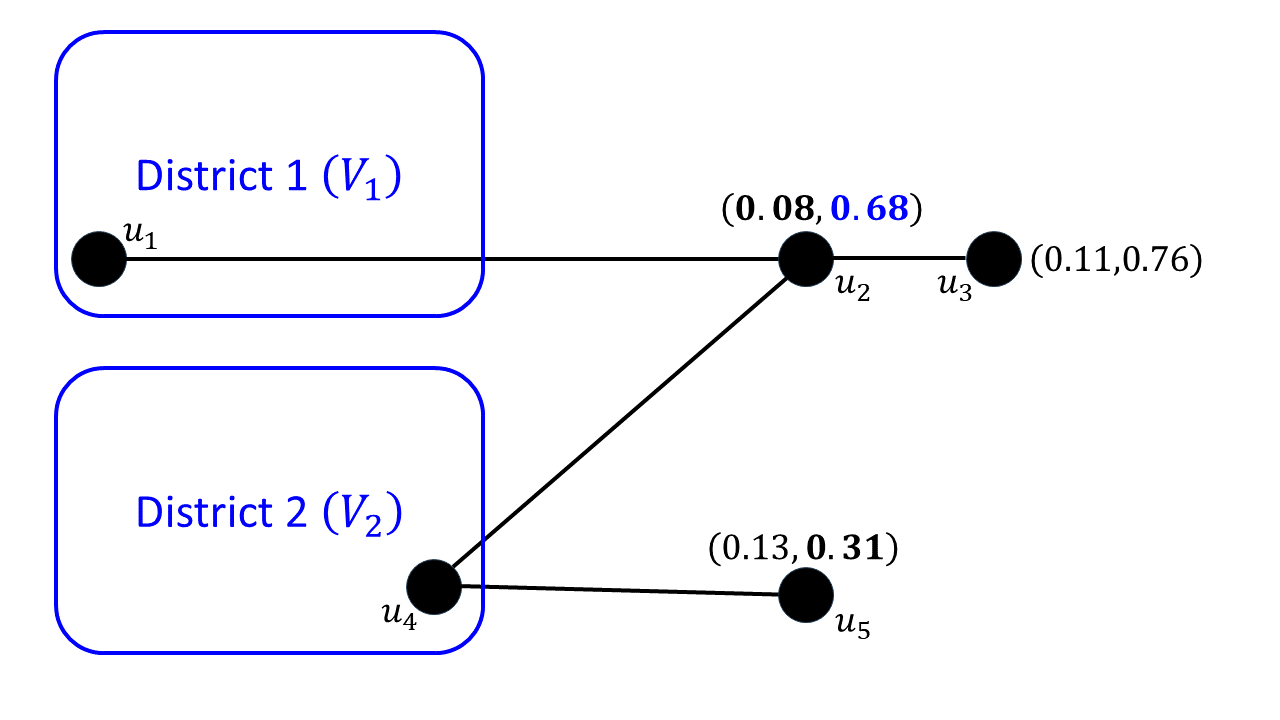}  
  \caption{Put $u_2$ into $V_2$ with {\bf $\tilde{Y}(u_2,2)$} larger than $\tilde{Y}(u_2,1),\tilde{Y}(u_5,2)$}
  \label{fig:sub-second}
\end{subfigure}
\newline
\begin{subfigure}{.5\textwidth}
  \centering
  % include third image
  \includegraphics[width=.8\linewidth]{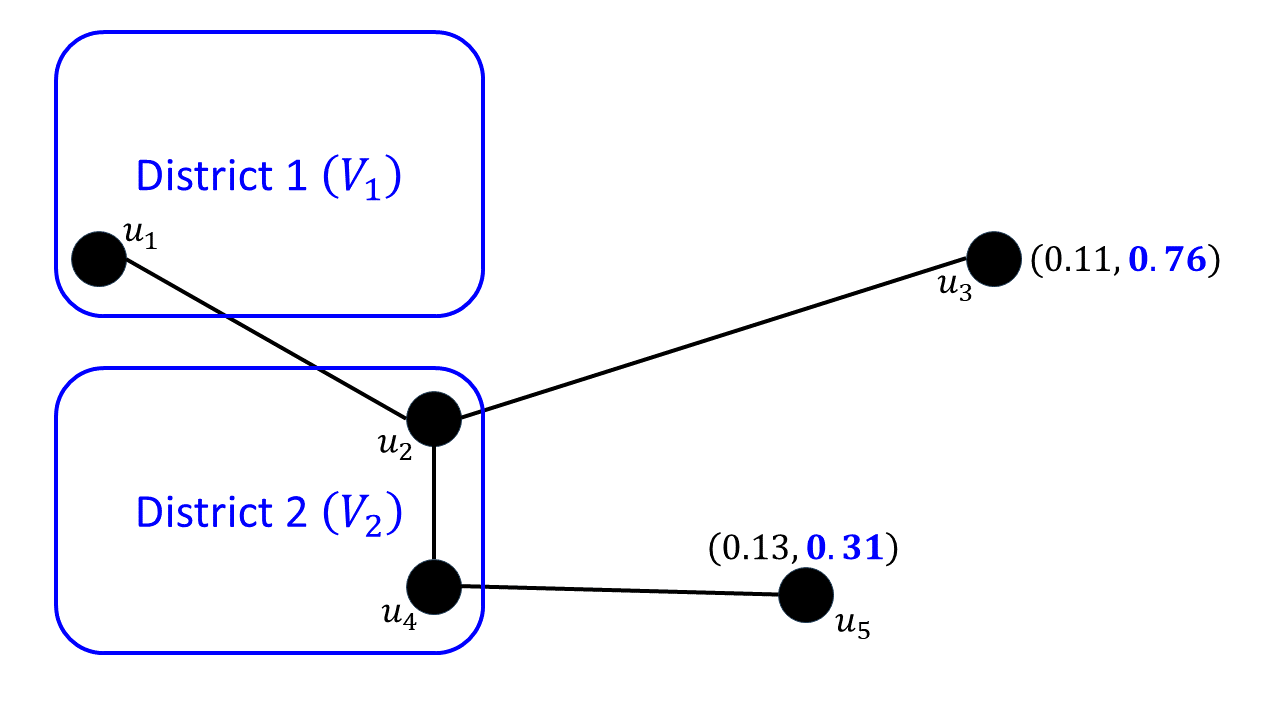}  
  \caption{Put $u_3,u_5$ into $V_2$ with their neighbors}
  \label{fig:sub-third}
\end{subfigure}
\begin{subfigure}{.5\textwidth}
  \centering
  % include fourth image
  \includegraphics[width=.8\linewidth]{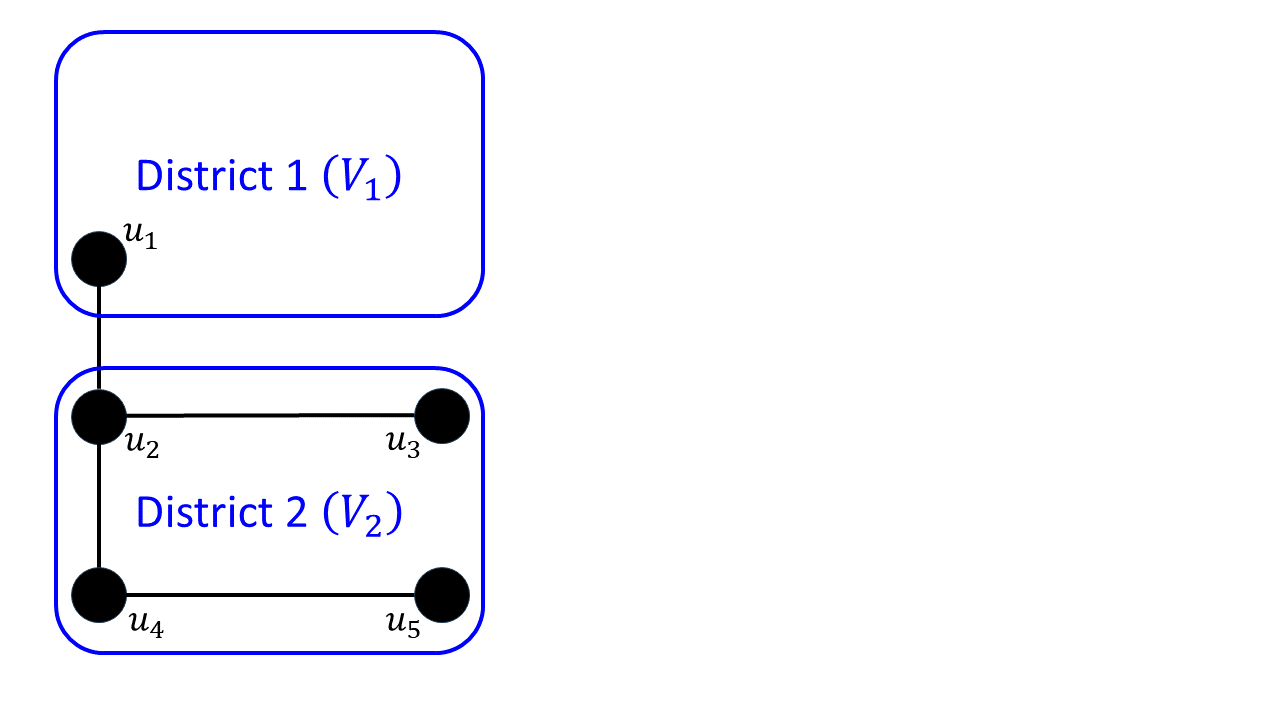}  
  \caption{$Y^{\mathrm{INT}}=\mathrm{ROUND}\left( \tilde{Y}\right)$}
  \label{fig:sub-fourth}
\end{subfigure}
\caption{Example of Rounding Algorithm}
\label{fig:fig}
\end{figure}

Figure~\ref{fig:fig} illustrate an example of implementation of Algorithm~\ref{a:rounding}. The tuple by each node $u_i$ is $\left( \tilde{Y}\left( u_i,1\right), \tilde{Y}\left( u_i,2\right) \right)$. In Figure~\ref{fig:sub-first}, $u_1$ and $u_4$ are to be put into the empty $V_1$ and $V_2$ with maximum values of $\tilde{Y}\left( u_1,1\right)$ and $\tilde{Y}\left( u_4,2\right)$, respectively (Lines~\ref{l:ifempty}-\ref{l:endifempty} of Algorithm~\ref{a:rounding}). In Figure~\ref{fig:sub-second}, $u_2$ has neighbors in $V_1$ and $V_2$ and $u_5$ has a neighbor in $V_2$. The value of $\tilde{Y}\left( u_2,2\right)=0.68$ is maximum among $\tilde{Y}\left( u_2,1\right)$, $\tilde{Y}\left( u_2,2\right)$ and $\tilde{Y}\left( u_5,2\right)$, and $u_2$ is to be put into $V_2$ (Lines~\ref{l:else}-\ref{l:endelse}).
Likewise, $u_3$ is to be put into $V_2$ and $u_5$ is to be put into $V_2$ in Figure~\ref{fig:sub-third}.
Figure~\ref{fig:sub-fourth} illustrates the resulting plan $Y^{\mathrm{INT}} = \mathrm{ROUND}\left( \tilde{Y}\right)$:
\begin{eqnarray*}
    Y^{\mathrm{INT}}\left( u_1,1\right) = Y^{\mathrm{INT}}\left( u_2,2\right) = Y^{\mathrm{INT}}\left( u_3,2\right) = Y^{\mathrm{INT}}\left( u_4,2\right) = Y^{\mathrm{INT}}\left( u_5,2\right) = 1
\end{eqnarray*}
where the other components of $Y^{\mathrm{INT}}$ are all zero-valued.

\subsubsection{Randomization}
Instead of the exact probability distribution of the traditional randomized rounding,
our randomized rounding procedure perturbs a given seed $\tilde{Y}$ randomly so that the perturbed seed $\tilde{Y}^{\mathrm{PTB}}$ may satisfy proportionality; {\it i.e.}, $P\left[ Y^{\mathrm{INT}(1)}_u = 1\right] < P\left[ Y^{\mathrm{INT}(2)}_u = 1\right]$ if $\tilde{Y}^{(1)}_u<\tilde{Y}^{(2)}_u$ and $\tilde{Y}^{(1)}_v =\tilde{Y}^{(2)}_v$ for $v\in V\setminus\{ u\}$, where $\tilde{Y}^{\mathrm{PTB}(i)}$ denote random perturbations of $\tilde{Y}^{(i)}$ and $Y^{\mathrm{INT}(i)} = \mathrm{ROUND}\left(\tilde{Y}^{\mathrm{PTB}(i)}\right)$ for $i=1,2$. We may consider two ways of perturbations:
\begin{eqnarray}
\tilde{Y}^{\mathrm{PTB}}_u &=& \tilde{Y}_u \tilde{\pi}_u\mbox{ with }\tilde{\pi}_u\in \mathrm{U}\left[ 0,1\right]\mbox{ for }u\in V,\label{e:perturb2}\\
\tilde{Y}^{\mathrm{PTB}}_u &=& \tilde{Y}_u + (1 - \tilde{Y}_u)\tilde{\pi}_u\mbox{ with }\tilde{\pi}_u\in \mathrm{U}\left[ 0,1\right]\mbox{ for }u\in V,\label{e:perturb}
\end{eqnarray}
which are uniform distributions $\mathrm{U}[0,\tilde{Y}_u]$ and $\mathrm{U}[\tilde{Y}_u,1]$ respectively.
Note that each way of perturbations (\ref{e:perturb2}) and (\ref{e:perturb}) satisfies both of (\ref{e:round}) and the proportionality.
In this paper, we use (\ref{e:perturb2}) for random perturbation.

\subsubsection{Moving Seed}
The initial seed for our randomized rounding procedure will be the centroid $\tilde{Y}^{(0)}=\dot{Y}=\left( \dot{Y}_u = 0.5 : u\in V\right)$ of $[0,1]^{V}\subset\mathbb{R}^{V}$, and the seed will move toward the best integer solution found so far, denoted by $Y^{\mathrm{BEST}}$, as better integer solutions are sought in the vicinity of $Y^{\mathrm{BEST}}$. The procedure first finds an integer solution $Y^{\mathrm{INT}(0)}=\mathrm{ROUND}\left( \tilde{Y}^{\mathrm{PTB}(0)}\right)$, where $\tilde{Y}^{\mathrm{PTB}(0)}$ is given by perturbing $\tilde{Y}^{(0)}=\dot{Y}$. The first integer solution is the best known integer solution $Y^{\mathrm{BEST}}=Y^{\mathrm{INT}(0)}$. At each trial $t\geq 1$, the procedure then moves the last seed $\tilde{Y}^{(t-1)}$, converging to the best known integer solution $Y^{\mathrm{BEST}}$ by exponential smoothing:
\begin{equation}
\tilde{Y}^{(t)} = \left( 1 - \alpha (t) \right)\tilde{Y}^{(t-1)} + \alpha (t) Y^{\mathrm{BKI}},\label{eqn:expSmoothNew}
\end{equation}
where $0<\alpha (t) < 1$. 
When the gerrymander score of $Y^{\mathrm{INT}(t)}$ is better than $Y^{\mathrm{BEST}}$ ({\it i.e.}, $S\left( Y^{\mathrm{BEST}}\right) > S\left( Y^{\mathrm{INT}(t)}\right)$), the procedure updates the best known integer solution with $Y^{\mathrm{BEST}} = Y^{\mathrm{INT}(t)}$.

\subsection{Full Procedure}\label{s:ARR}
In this section, we review the full adaptive randomized rounding (ARR) procedure which Kim and Shim\cite{KS2022} introduced. Converging to the best known integer solution $Y^{\mathrm{BEST}}$, the seed finds a better integer solution near $Y^{\mathrm{BEST}}$. However, if the seed is too close to the best known integer solution, it will stick to the best known integer solution finding only the same integer solution. To resolve this issue, the ARR adopts two criteria:
\begin{itemize}
    \item Criterion~1. Control of Speed: The seed $\tilde{Y}$ slows down near the best known integer solution.
    \item Criterion~2. Rule of Reset: If the seed is too close, reset $\tilde{Y}=\dot{Y}$.
\end{itemize}

During the convergence to the best known integer solution, a decelerator slows the speed of convergence (Criterion~1). To determine $\alpha (t)$, we define the root-mean-square deviation (RMSD) of $\tilde{Y}^{(t)}$ with respect to the centroid $\dot{Y}$, {\it i.e.},
\begin{equation}
    \mathrm{RMSD}(\tilde{Y}^{(t)}) = \sqrt{\frac{\sum_{(u,d)\in V\times D} \left(\tilde{Y}^{(t)}(u,d)-0.5\right)^2}{|V\times D|}},
\end{equation}
which ranges from 0 at $\dot{Y}$ to 0.5 at an integer solution $Y^{\mathrm{INT}}$. Thus, as $\tilde{Y}^{(t)}$ converges to the best known integer solution $Y^{\mathrm{BEST}}$, the RMSD converges to 0.5.
The smoothing constant $\alpha (t)$ is determined by decelerator (DEC) which slows $\alpha = \frac{1}{2}$ (at $\dot{Y}$) to $\alpha = \frac{1}{1+e^2}$ (at $Y^{\mathrm{BEST}}$):
\begin{equation}
    \label{e:decRR}
    \alpha (t) = \mathrm{DEC} (t-1) = \frac{1}{1+e^{4*\mathrm{RMSD}(\tilde{Y}^{(t-1)})}}.
\end{equation}

\begin{figure}
\centering
\includegraphics[width=0.9\textwidth]{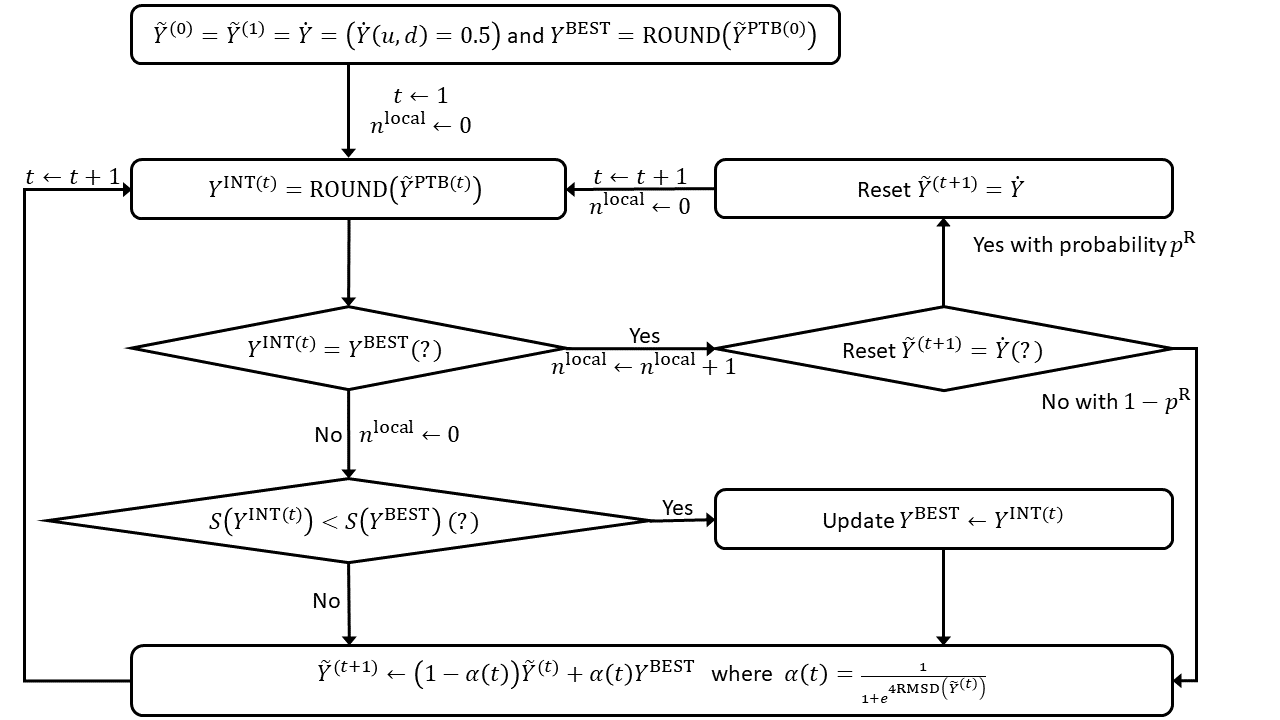}
\caption{Flowchart of ARR where the root-mean-square deviation (RMSD) of $\tilde{Y}$ with respect to 0.5 is $\mathrm{RMSD}\left(\tilde{Y}\right)=\sqrt{\frac{\sum_{u\in V}\sum_{d\in D}\left(\tilde{Y}(u,d) - 0.5\right)^2}{|V\times D|}}$ and the probability of reset is $p^\mathrm{R}=\min\left(\frac{n^\mathrm{local}}{20}, 1\right) *\mathrm{RMSD}\left( \tilde{Y}^{(t)} \right)$.}\label{f:ARR}
\end{figure}

This ARR procedure is illustrated by the flowchart in Figure~\ref{f:ARR}. 
The number of consecutive trials in which the same solution $Y^{\mathrm{BEST}}$ is found is denoted by $n^{\mathrm{local}}$ (for Criterion~2). 
As the same solution $Y^{\mathrm{BEST}}$ is found in the run of trials ({\it i.e.}, $n^{\mathrm{local}}$ increases), the probability that the seed will be reset to $\dot{Y}$ increases. In particular, the probability of reset $p^\mathrm{R}$ reaches $\mathrm{RMSD}\left(\tilde{Y}^{(t)}\right)$ after a run of 20 trials that find the same $Y^{\mathrm{BEST}}$.
The maximum number of trials may be set as the termination criterion for the full procedure. 

\subsection{Two Phases}\label{s:phase1}
Let $M$ be a big number larger than any gerrymander score; {\it e.g.}, $M = 1 + \sum_{v\in V}s_{uv}$ for one node $u\in V$.
To identify one feasible solution $Y^{\mathrm{INT}}$ satisfying the equal population constraints, Phase~I problem minimizes the infeasibility score:
\begin{eqnarray}
    S_{\mathrm{I}}\left( Y^{\mathrm{INT}}\right) = M + \sum_{d\in D}\max\left( 0,- p_{\max} + \sum_{u\in V}p(u)Y^{\mathrm{INT}}(u,d) \right) + \max\left( 0, p_{\min} - \sum_{u\in V}p(u)Y^{\mathrm{INT}}(u,d) \right).
\end{eqnarray}
Our ARR procedure may set a conditional objective function:
\begin{eqnarray}
    S\left( Y^{\mathrm{INT}}\right) =
    \left\{
    \begin{array}{l}
    S_{\mathrm{I}}\left( Y^{\mathrm{INT}}\right)\mbox{ if }Y^{\mathrm{INT}}\mbox{ is infeasible}\\
    S_{\mathrm{II}}\left( Y^{\mathrm{INT}}\right)\mbox{ if }Y^{\mathrm{INT}}\mbox{ is feasible}
    \end{array}
    \right\}
\end{eqnarray}
where $S_{\mathrm{II}}\left( Y^{\mathrm{INT}}\right)$ is the gerrymander score of the district plan indicated by $Y^{\mathrm{INT}}$.

\section{The Case Study: 2012 Maryland congressional district plan}\label{s:case}
Following a plan proposed by the Democrats, Maryland’s 2012 electoral lines were redrawn in 2011. However, there was a large amount of controversy surrounding this map.~\cite{lwv2019} In particular the sixth congressional district drew a large amount of scrutiny from Republican critics; this district was seen as a case of cracking Republican support.~\cite{bailey2020} The cracking of Republic support severely weakened the Republican party's political power in the state as it gave the Democrats one more seat in Congress.~\cite{portnoy2019} Critics were quick to point out this fact and eventually the case was taken to the Supreme Court. Governor Lawrence J. Hogan Jr. submitted an amicus brief supporting plaintiffs that challenged the Democrats' electoral map.~\cite{wjz2019} Despite his lobbying, the Supreme Court threw out his case stating that the Supreme Court and other courts have no place in settling partisan gerrymandering claims.~\cite{kurtz2019} By ruling that federal courts could not interfere in the electoral maps of each state, the Supreme Court set the precedent that gerrymandering could go unpunished. As stated by Governor Hogan, ``(the court’s ruling) terribly disappointing to all who believe in fair elections.''~\cite{portnoy2019}

\begin{figure}
\centering
\includegraphics[width=0.9\textwidth]{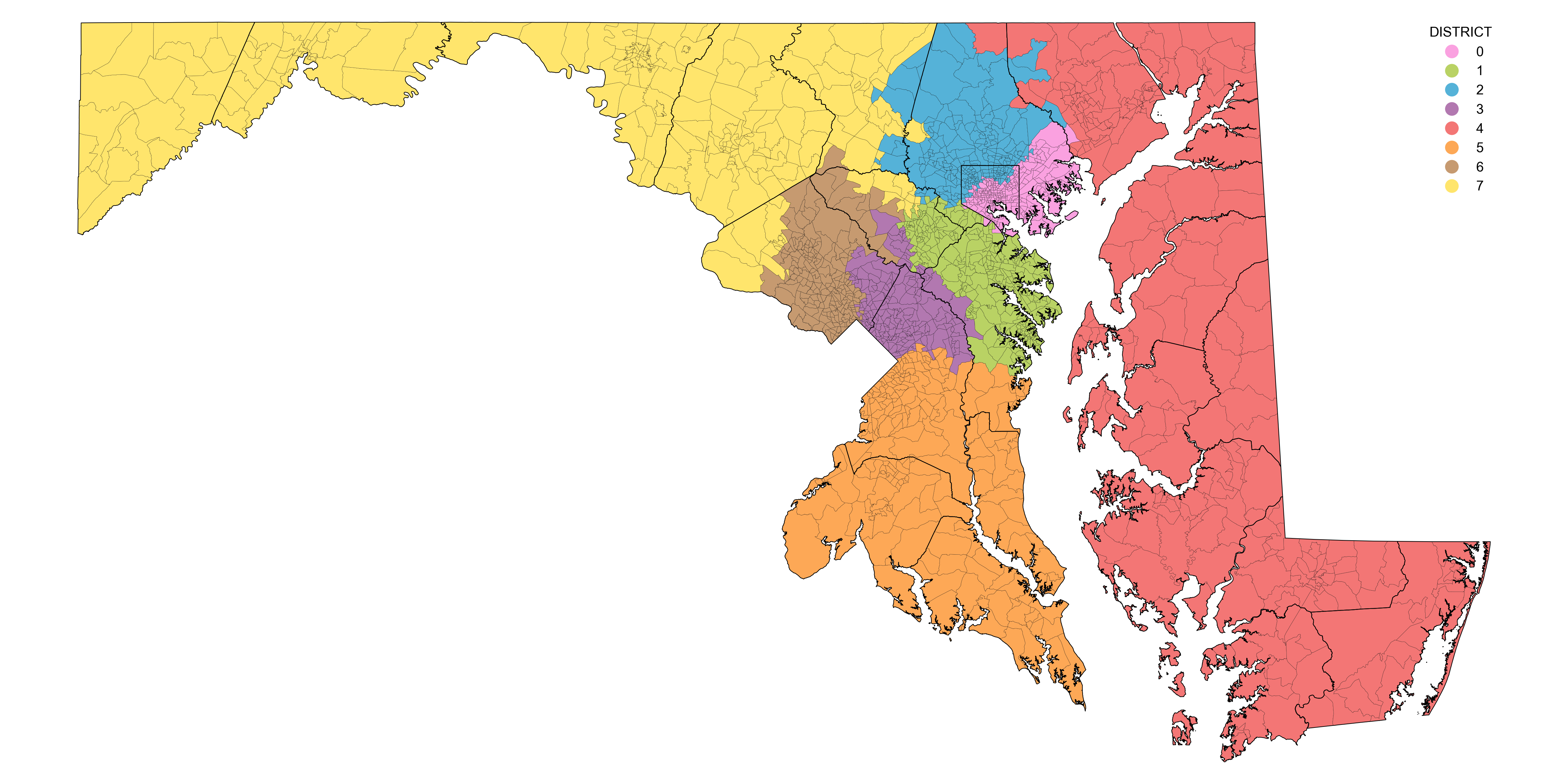}
\caption{Redrawing the 2012 map of Maryland congressional districts (gerrymander score = 10,152)}\label{f:newMap}
\end{figure}

Chopra, Park and Shim~\cite{chopra2023} redrew the map partitioning 46 coarse population units into the 8 congressional districts. However, the real-world political districting problem partitions 1,849 precincts in Maryland. Using our adaptive randomized rounding (ARR) heuristic (Section~\ref{s:heuristic}), a usual digital computer solved the large-scale political districting problem to redistrict the 1,849 precincts of Maryland. 
In Phase~I, ARR identified a feasible connected partition $Y^{\mathrm{INT}}$ satisfying all the equal population constraints (\ref{e:population}). 
Then, Phase~II improved the feasible solution to a good solution of a small gerrymander score 10,152 (distance from a central precinct $\approx 5.5$ on average).
Figure~\ref{f:newMap} is the good solution. 
The fictitious districts are numbered from 0 to 7. 
In the figure, we see the shape of the fictitious District~7 is very different from the sixth district~\footnote{See the actual 2012 map at \url{https://planning.maryland.gov/Redistricting/Pages/2010/congDist.aspx}} which raised the gerrymander issue in 2019. 

\section{Conclusion}\label{s:conclusion}
Chopra, Park and Shim~\cite{chopra2023} developed an integer linear program (ILP) of the political districting problem which Mehrotra, Johnson and Nemhauser~\cite{mehrotra1998optimization} had introduced and redrew the map partitioning 46 coarse population units into the 8 congressional districts. Integrating the ILPs developed by Hess et~al.~\cite{hess1965} and Shirabe~\cite{shirabe2009}, Validi, Buchanan and Lykhovyd~\cite{validi2022} introduced a very efficient ILP. However, the real-world political districting problem partitions 1,849 precincts\footnote{\url{https://planning.maryland.gov/Redistricting/Pages/2010/precinct.aspx}} in Maryland.\footnote{The adjacency graph consists of total 5,310 edges. For connectivity, we added to the adjacency graph the nine more edges in Table~\ref{t:addEdge}.} The efficient ILP cannot solve even the root simplex using supercomputers\cite{center1987ohio,towns2014xsede} for 50 hours. Our adaptive randomized rounding algorithm produced a good district plan in which the shape of the sixth district is clearly different from that in the original 2012 map while the original shape was same as that in the optimal plan given by Chopra et~al.~\cite{chopra2023}.

\begin{table} 
\caption{Additional edges of the adjacency graph}\label{t:addEdge}
\begin{center}
\begin{tabular}{| c | c | c |}
\hline
Edge No.	&Source	&Target\\
\hline
5311	&St. Mary's Precinct 02-001	&St. Mary's Precinct 09-001\\
5312	&St. Mary's Precinct 09-001	&St. Mary's Precinct 02-002\\
5313	&Somerset Precinct 10-001	&Somerset Precinct 10-002\\
5314	&Somerset Precinct 14-001	&Somerset Precinct 10-002\\
5315	&Somerset Precinct 14-001	&Somerset Precinct 09-001\\
5316	&Worcester Precinct 07-001	&Worcester Precinct 03-002\\
5317	&Worcester Precinct 07-001	&Worcester Precinct 03-001\\
5318	&Worcester Precinct 07-001	&Worcester Precinct 06-003\\
5319	&Cecil Precinct 02-001	    &Cecil Precinct 02-002\\
\hline
\end{tabular}
\end{center}
\end{table}

\end{document}